\documentclass[%
 reprint,
 pre,
amsmath,amssymb,
aps,
]{revtex4-1}
\usepackage[abs]{overpic}
\usepackage{graphicx}% Include figure files
\usepackage{dcolumn}% Align table columns on decimal point
\usepackage{bm}% bold math
\usepackage{placeins}
\begin{document}

\preprint{APS/123-QED}

\title{Mechanisms for the clustering of inertial particles in the inertial range of isotropic turbulence}% Force line breaks with \\

\author{Andrew D. Bragg}
 \email{adb265@cornell.edu}
\author{Peter J. Ireland}
\author{Lance R. Collins}%
\affiliation{%
Sibley School of Mechanical \& Aerospace Engineering, Cornell University, Ithaca, NY 14853}%

%\affiliation{
%\affiliation{
%\affiliation{%

\date{\today}% It is always \today, today,

\begin{abstract}

In this paper, we consider the physical mechanism for the clustering of inertial particles in the inertial range of isotropic turbulence.  We analyze the exact, but unclosed, equation governing the radial distribution function (RDF) and compare the mechanisms it describes for clustering in the dissipation and inertial ranges.  We demonstrate that in the limit ${St_r\ll1}$, where $St_r$ is the Stokes number based on the eddy turnover timescale at separation $r$, the clustering in the inertial range can be understood to be due to the preferential sampling of the coarse-grained fluid velocity gradient tensor at that scale.  When ${St_r\gtrsim\mathcal{O}(1)}$ this mechanism gives way to a non-local clustering mechanism.  These findings reveal that the clustering mechanisms in the inertial range are analogous to the mechanisms that we identified for the dissipation regime (see \emph{New J. Phys.} \textbf{16}:055013, 2014). Further, we discuss the similarities and differences between the clustering mechanisms we identify in the inertial range and the ``sweep-stick'' mechanism developed by Coleman \& Vassilicos (Phys. Fluids 21:113301, 2009). We argue that when ${St_r\ll1}$ the sweep-stick mechanism is equivalent to our mechanism in the inertial range if the particles are suspended in Navier-Stokes turbulence, but that the sweep-stick mechanism breaks down for ${St_r\gtrsim\mathcal{O}(1)}$. The argument also explains why the sweep-stick mechanism is unable to predict particle clustering in kinematic simulations. We then consider the closed, model equation for the RDF given in Zaichik \& Alipchenkov (Phys. Fluids. 19:113308, 2007) and use this, together with the results from our analysis, to predict the analytic form of the RDF in the inertial range for ${St_r\ll1}$, which, unlike that in the dissipation range, is not scale-invariant.  The results are in good agreement with direct numerical simulations, provided the separations are well within the inertial range.

%\begin{description}
%\end{description}
\end{abstract}

\maketitle

%\tableofcontents

%\section{First section}
%
%Article content.
%
%\subsection{A Subsection}

\section{Introduction}

An initially uniform distribution of inertial particles in an incompressible turbulent fluid velocity field will develop dynamically evolving spatial clusters.  Such clustering has important implications for aerosol processes such as gravitational settling~\cite{maxey87,wang93}, turbulence modulation~\cite{elghobashi93,sundaram6} and particle collisions~\cite{sundaram4,wwz00}.  These processes are relevant to industrial processes such as aerosol manufacturing~\cite{moody03}, drug delivery~\cite{li96} and spray combustion~\cite{faeth96} as well as to natural processes such as sediment and plankton distribution in oceans~\cite{malkiel06} and even the formation of planets in the early universe~\cite{johansen07}. 

In a recent paper \cite{bragg14b}, we considered in detail the physical mechanism responsible for the clustering of inertial particles in the dissipation range of isotropic turbulence.  Formally, the dissipation range is defined as ${r\ll\eta}$, where $r$ is the distance between two points in space and $\eta$ is the Kolmogorov length scale, though it should be noted that experiments and numerical simulations of the Navier-Stokes equation suggest that the dissipation range actually extends to $r=\mathcal{O}(10\eta)$ \cite{ishihara09}.  Nevertheless, in what follows we define the dissipation range to be the limit ${r\ll\eta}$. In \cite{bragg14b} we showed that in the limit ${St\ll1}$ (where $St\equiv\tau_p/\tau_\eta$ is the Stokes number, $\tau_p$ is the particle response time and $\tau_\eta$ is the Kolmogorov timescale), the mechanism for clustering in the Zaichik \& Alipchenkov theory \cite{zaichik03,zaichik07,zaichik09} (hereafter this body of work is referred to as `ZT') is the same as that in the Chun \emph{et al.} theory \cite{chun05} (hereafter referred to as `CT'), which is essentially an extension of the classical argument of Maxey \cite{maxey87} that particles are centrifuged out of rotating regions of the fluid into regions of high strain rate.  When ${St\gtrsim\mathcal{O}(1)}$, we showed that the ZT describes an additional non-local contribution to the clustering mechanism that is discussed in greater detail in \S\ref{ACMIR}.  

If the Taylor microscale Reynolds number, $Re_\lambda$, is sufficiently large, particles may also cluster in the inertial range of the turbulence, a scenario that has been considered in several works \cite{bec07,bec08,chen06,goto06,goto08,coleman09,pan11}.  The inertial range is defined as ${\eta\ll r\ll L}$, where $L$ is the integral lengthscale of the turbulence.  In \cite{bec07}, they showed using direct numerical simulations (DNS) that particle clustering at ${\eta\ll r\ll L}$ is not scale-invariant, unlike for ${r\ll\eta}$. Furthermore, they argued that the clustering is not simply characterized by $St_r$, as would be predicted by a white-in-time flow analysis (e.g. \cite{bec08}), but rather by a rescaled contraction rate, at least for ${St_r\ll1}$, where $St_r\equiv\tau_p/\langle\epsilon\rangle^{-1/3} r^{2/3}$ is the scale-dependent particle Stokes number based on eddies of size $r$, and $\langle\epsilon\rangle$ is the average turbulent energy dissipation rate.  In a series of articles \cite{chen06,goto06,goto08,coleman09}, an explanation for clustering at ${\eta\ll r\ll L}$ was developed in terms of the ``sweep-stick'' mechanism, whereby inertial particles are argued to stick to stagnation points in the fluid acceleration field and are swept along with them by the local fluid velocity.  Since the fluid acceleration stagnation points are clustered in Navier-Stokes turbulence, they argue that this leads to clustering of the inertial particles at ${\eta\ll r\ll L}$. Moreover, in \cite{coleman09}, they argue that the clustering mechanisms operating at ${r\ll\eta}$ and ${\eta\ll r\ll L}$ are different, with the sweep-stick mechanism describing the clustering only for ${\eta\ll r\ll L}$.  The break in scale-invariance of the clustering noted in \cite{bec07} as one goes from the dissipation range to the inertial range is certainly consistent with their hypothesis of different clustering mechanisms operating in the two regimes.

The outline of the paper is as follows.  In \S\ref{ACMIR} we examine the question of the clustering mechanism in the inertial range by analyzing the exact equation for the radial distribution function (RDF), and show that the mechanism is precisely analogous to that operating in the dissipation range.  We show that the break in scale-invariance of the clustering does not arise from a change in the underlying mechanism. In \S\ref{SSmech}, we contrast our findings with the sweep-stick model of Coleman \& Vassilicos \cite{coleman09}. Finally, in \S\ref{RDF_pred} we apply our findings to the model equation for the RDF from Zaichik \& Alipchenkov \cite{zaichik07} and derive a prediction for the analytical form of the RDF in the inertial range for ${St_r\ll1}$, which we test against DNS data at ${Re_\lambda=597}$.  

\section{Analysis of the clustering mechanism in the inertial range}\label{ACMIR}

We consider the relative motion between two identical point particles, a `primary' particle and a `satellite' particle. We make the approximations that the particles are subject to Stokes drag forces only, that they do not interact with each other through physical collisions or hydrodynamic interactions and that they are at low enough concentration to not affect the turbulence (i.e., `one-way coupling'). Furthermore, we restrict our attention to statistically stationary, homogeneous and isotropic turbulence. One of the reasons for choosing such simplified turbulence and particle dynamics is that we want to compare our analysis with earlier studies that were based on the same simplifications \citep[e.g.][]{bec07,bec08,chen06,goto06,goto08,coleman09,pan11}.  The equation governing the relative motion of the two particles is \citep{maxey83}
\begin{equation}
\dot{\boldsymbol{w}}^p(t)=(St\tau_{\eta})^{-1}\Big(\Delta\boldsymbol{u}(\boldsymbol{r}^p(t),t)-\boldsymbol{w}^p(t)\Big),\label{eom}	
\end{equation}
where $\boldsymbol{r}^p(t),\boldsymbol{w}^p(t), \dot{\boldsymbol{w}}^p(t)$ are the particle pair relative separation, relative velocity and relative acceleration vectors, respectively, and $\Delta\boldsymbol{u}(\boldsymbol{r}^p(t),t)$ is the difference in the fluid velocity evaluated at the positions of the two particles.  

For the system governed by (\ref{eom}) the exact equation governing the probability density function (PDF) $p(\boldsymbol{r},\boldsymbol{w},t)\equiv\langle\delta(\boldsymbol{r}^p(t)-\boldsymbol{r})\delta(\boldsymbol{w}^p(t)-\boldsymbol{w})\rangle$ describing the distribution of $\boldsymbol{r}^p(t),\boldsymbol{w}^p(t)$ in the phase-space $\boldsymbol{r},\boldsymbol{w}$  is
\begin{align}
\begin{split}
\partial_t p=&-\boldsymbol{\nabla_r\cdot}p\boldsymbol{w}+(St\tau_{\eta})^{-1}\boldsymbol{\nabla_w\cdot}p\boldsymbol{w}\\
&-(St\tau_{\eta})^{-1}\boldsymbol{\nabla_w\cdot}p\langle\Delta\boldsymbol{u}(\boldsymbol{r}^p(t),t)\rangle_{\boldsymbol{r},\boldsymbol{w}},\label{PDFeq}
\end{split}
\end{align}
where $\langle\cdot\rangle_{\boldsymbol{r},\boldsymbol{w}}$ denotes an ensemble average conditioned on ${\boldsymbol{r}^p(t)=\boldsymbol{r}}$ and ${\boldsymbol{w}^p(t)=\boldsymbol{w}}$.  
A commonly used statistical measure of particle clustering is the RDF \cite{mcquarrie}, which is defined as the ratio of the number of particle pairs at separation ${r=\vert\boldsymbol{r}\vert}$ to the number that would be expected if the particles were uniformly distributed.  An exact equation for the statistically stationary RDF, $g(\boldsymbol{r})$, can be constructed by multiplying the stationary form of (\ref{PDFeq}) by $\boldsymbol{w}$ and then integrating over all $\boldsymbol{w}$ yielding
\begin{equation}
\boldsymbol{0}=g\langle\Delta\boldsymbol{u}(\boldsymbol{r}^p(t),t)\rangle_{\boldsymbol{r}}-St\tau_{\eta}\boldsymbol{S}^{p}_{2}\boldsymbol{\cdot\nabla_r}g-St\tau_{\eta}g\boldsymbol{\nabla_r\cdot}\boldsymbol{S}^{p}_{2},\label{RDFeq}
\end{equation}
where 
\begin{align}
g(\boldsymbol{r})=\frac{N(N-1)}{n^2 V}\int_{\boldsymbol{w}} p(\boldsymbol{r},\boldsymbol{w})\,d\boldsymbol{w},	
\end{align}
$N$ is the total number of particles lying within the control volume $V$, ${n\equiv N/V}$ is the number density of particles, and ${\boldsymbol{S}^{p}_{2}(\boldsymbol{r})\equiv\langle\boldsymbol{w}^p(t)\boldsymbol{w}^p(t)\rangle_{\boldsymbol{r}}}$ is the second-order particle velocity structure function.

The drift mechanisms that generate clustering are associated with the term ${St\tau_{\eta}\bm{\nabla_r\cdot}\bm{S}^{p}_{2}}$.  The contribution from ${g\langle\Delta\boldsymbol{u}(\boldsymbol{r}^p(t),t)\rangle_{\boldsymbol{r}}}$ may also contain drift contributions in addition to diffusion effects (see \cite{bragg14b}), and this term is unclosed. It is not necessary at this stage to consider closure approximations for ${g\langle\Delta\boldsymbol{u}(\boldsymbol{r}^p(t),t)\rangle_{\boldsymbol{r}}}$ since its physical interpretation is known, namely it describes a flux arising from correlations between $\Delta\boldsymbol{u}$ and $\bm{r}^p(t)$ that is  associated with preferential sampling effects.  Hence for this qualitative discussion, we will focus on understanding the physical mechanisms described by the term ${St\tau_{\eta}\bm{\nabla_r\cdot}\bm{S}^{p}_{2}}$.

We begin by reviewing the findings from \cite{bragg14b} on the meaning and behavior of ${St\tau_{\eta}\bm{\nabla_r\cdot}\bm{S}^{p}_{2}}$ in the dissipation range.
In \cite{bragg14b} we showed that for ${r\ll\eta}$ and $St\ll1$
\begin{align}
St\tau_{\eta}\bm{\nabla_r\cdot}\bm{S}^{p}_{2}\approx\frac{St\tau_\eta}{3}\bm{r}(\mathcal{A}-\mathcal{B}),\label{ZTlowSt}
\end{align}
where ${\mathcal{A}\equiv\langle\mathcal{S}^2(\bm{x}^p(t),t)\rangle}$ and ${\mathcal{B}\equiv\langle\mathcal{R}^2(\bm{x}^p(t),t)\rangle}$ are averages of the second invariants of the strain-rate $\bm{\mathcal{S}}$ and rotation-rate $\bm{\mathcal{R}}$ tensors evaluated along the inertial particle trajectory $\bm{x}^p(t)$.  This drift mechanism is identical to the one derived in the CT using perturbation theory, and is associated with the traditional centrifuge mechanism.  For ${St\gtrsim\mathcal{O}(1)}$, the particle velocity dynamics become increasingly non-local, and this fundamentally changes the clustering mechanism described by $St\tau_{\eta}\bm{\nabla_r\cdot}\bm{S}^{p}_{2}$. The physical interpretation of the non-local drift is as follows. Particle pairs arriving at separation $\bm{r}$ coming from larger separations carry a memory of larger fluid velocity differences in their path-history as compared with pairs arriving at $\bm{r}$ from smaller separations.  This path-history bias breaks the symmetry of the particle inward and outward motions, creating a net inward drift and clustering. 

In order to analyze the clustering mechanism in the inertial range, we consider the limit ${Re_\lambda\rightarrow\infty}$, such that the inertial range is unbounded. Furthermore, we define a scale-dependent Stokes number as ${St_r\equiv\tau_p/\tau_r}$, where $\tau_r$ is the eddy turnover timescale defined as ${\tau_r\sim\langle\epsilon\rangle^{-1/3}r^{2/3}}$ for ${\eta\ll r\ll L}$, $\langle\epsilon\rangle$ is the average turbulent kinetic energy dissipation rate and $L$ is the (asymptotically large) integral length scale. For arbitrary Stokes numbers, $St$, the limit ${St_r\ll 1}$ corresponds to ${r\gg\eta St^{3/2}}$. We can analyze this regime in much the same way as CT did for ${r\ll\eta}$ and ${St\ll1}$.

Introducing the coarse-grained strain-rate $\widetilde{\bm{\mathcal{S}}}$ and rotation-rate $\widetilde{\bm{\mathcal{R}}}$ tensors, with coarse-graining length scale $r$, we can write the fluid velocity difference as ${\Delta\bm{u}(\bm{r},t)\sim(\widetilde{\bm{\mathcal{S}}}+\widetilde{\bm{\mathcal{R}}})\bm{\cdot}\bm{r}}$ \cite{naso05,li05,li10}. In the limit ${St_r\ll1}$, ${\bm{w}^p(t)\approx\Delta\bm{u}(\bm{r}^p(t),t)+\mathcal{O}(St_r)}$ and therefore to leading order ${St\tau_{\eta}\bm{\nabla_r\cdot}\bm{S}^{p}_{2}}$ is ${St\tau_{\eta}\bm{\nabla_r\cdot}\langle\Delta\bm{u}(\bm{r}^p(t),t)\Delta\bm{u}(\bm{r}^p(t),t)\rangle_{\bm{r}}}$.  We can derive an expression for the latter quantity using the coarse-graining and the scaling from Kolmogorov's 1941 theory (K41, see \cite{frisch}), yielding
\begin{align}
%\begin{split}
St\tau_{\eta}\bm{\nabla_r\cdot}\bm{S}^{p}_{2}\approx
\frac{St\tau_\eta}{3}\bm{r}\Big[\frac{2r}{5}\nabla_r\widetilde{\mathcal{A}}+\widetilde{\mathcal{A}}-\zeta\widetilde{\mathcal{B}}\Big]\label{DVir},
%\end{split}
\end{align}
where $\widetilde{\mathcal{A}}\equiv\langle\widetilde{\bm{\mathcal{S}}^p}\bm{:}\widetilde{\bm{\mathcal{S}}^p}\rangle$, $\widetilde{\mathcal{B}}\equiv\langle\widetilde{\bm{\mathcal{R}}^p}\bm{:}\widetilde{\bm{\mathcal{R}}^p}\rangle$, $\widetilde{\bm{\mathcal{S}}^p}$ and $\widetilde{\bm{\mathcal{R}}^p}$ denote $\bm{\mathcal{S}}(\bm{x}^p(t),t)$ and $\bm{\mathcal{R}}(\bm{x}^p(t),t)$ coarse-grained over the scale $r$, $\zeta(r\ll\eta)=1$ and $\zeta(\eta\ll r\ll L)=7/15$ \footnote[1]{The factor $\zeta(r)$ arises because in using the coarse-grained approximation one must account for the statistical dependence of $\Delta\bm{u}(\bm{r},t)$ on the orientation of $\bm{r}$.}.  For ${\eta\ll r\ll L}$ (\ref{DVir}) becomes
\begin{align}
St\tau_{\eta}\bm{\nabla_r\cdot}\bm{S}^{p}_{2}=\frac{7 St\tau_\eta}{45}\bm{r}(\widetilde{\mathcal{A}}-\widetilde{\mathcal{B}}),
\label{eq:drift}
\end{align}
and for ${r\ll\eta}$, (\ref{DVir}) reduces to (\ref{ZTlowSt}). Preferential sampling of the inertial range eddies will lead to $\widetilde{\mathcal{A}}>\widetilde{\mathcal{B}}$, which is associated with centrifuging out of eddies of size $\sim r$.  Note that any drift contribution coming from the unclosed term $\langle\Delta\boldsymbol{u}(\boldsymbol{r}^p(t),t)\rangle_{\boldsymbol{r}}$ in (\ref{RDFeq}) has a similar interpretation.  

At separations ${r\lesssim\mathcal{O}(\eta St^{3/2})}$, corresponding to ${St_r\gtrsim\mathcal{O}(1)}$, so long as $\Delta\bm{u}(\bm{r},t)$ is statistically dependent upon $\bm{r}$, the non-local, path-history symmetry breaking contribution to ${St\tau_{\eta}\bm{\nabla_r\cdot}\bm{S}^{p}_{2}}$ becomes important. This transition is analogous to the one that occurs in the dissipation range (i.e., ${r\ll\eta}$) for particles with ${St\gtrsim\mathcal{O}(1)}$. However, the relative magnitude of the transition from the local to the non-local mechanisms is more pronounced in the dissipation range than in the inertial range.
%
%It is important to emphasize that whereas the non-local clustering mechanism dominates at ${r\ll\eta}$ when ${St\gtrsim\mathcal{O}(1)}$ (see \cite{bragg14b}), the same may not be true for ${\eta\ll r\ll L}$ when ${St_r\gtrsim\mathcal{O}(1)}$.
The reason for this is that, although the particle relative velocities have a non-local contribution when ${St_r\gtrsim\mathcal{O}(1)}$, the non-locality is much weaker in the inertial range than in the dissipation range because the dependence of $\Delta\bm{u}(\bm{r},t)$ on $\bm{r}$ is weaker in the inertial range. Consequently,
% at ${St_r\gtrsim\mathcal{O}(1)}$
the filtering effect of the particle inertia (see \cite{salazar12a}) can dominate the non-local contribution to the particle relative velocities leading to ${\bm{S}^p_2/\langle\Delta\bm{u}(\bm{r},t)\Delta\bm{u}(\bm{r},t)\rangle<1}$.  DNS results show that whereas ${\bm{S}^p_2/\langle\Delta\bm{u}(\bm{r},t)\Delta\bm{u}(\bm{r},t)\rangle\gg1}$ for ${St\gtrsim\mathcal{O}(1)}$ in the dissipation range, ${\bm{S}^p_2/\langle\Delta\bm{u}(\bm{r},t)\Delta\bm{u}(\bm{r},t)\rangle<1}$ for ${St_r\gtrsim\mathcal{O}(1)}$ in the inertial range~\cite{ireland14}. However, the latter result is sensitive to the Reynolds number. In particular, in the limit ${Re_\lambda\to\infty}$, where the filtering effect of particle inertia on the largest scales of the flow vanishes, the non-local clustering mechanism dominates the inertial range for ${St_r\gtrsim\mathcal{O}(1)}$. 

We therefore conclude that the clustering mechanisms operating in the inertial range are analogous to those operating in the dissipation range, with the coarse-grained strain and rotation in the inertial range playing the role of the strain and rotation in the dissipation range. When $St_r\ll1$ preferential sampling of the coarse-grained fluid velocity gradient tensor at scale $\sim r$ generates the inward drift and clustering, and when ${St_r\gtrsim\mathcal{O}(1)}$ the non-local, path-history symmetry breaking mechanism contributes to the clustering.

\section{Relationship to the sweep-stick mechanism}\label{SSmech}

%In the previous section we have argued that the mechanism generating clustering at ${r\ll\eta}$ and ${\eta\ll r\ll L}$ is precisely analogous.  This is in contradiction to the arguments in \cite{goto06,chen06,goto08,coleman09} where it is argued that a completely different mechanism generates clustering when ${\eta\ll r\ll L}$, namely the ``sweep-stick'' mechanism.

As noted earlier, there is an alternative description of inertial particle clustering known as the ``sweep-stick'' mechanism~\cite{goto06,chen06,goto08,coleman09}.
The sweep-stick mechanism was motivated by the observation that the instantaneous particle positions $\bm{x}^p(t)$ are correlated with the positions of the stagnation points of the acceleration field of the fluid, $\bm{s}_a(t)$, defined such that $\bm{a}(\bm{s}_a(t),t)\equiv\bm{0}$. Chen \emph{et al.}~\cite{chen06} used K41 scaling to obtain
\begin{align}
\Big\langle\vert\dot{\bm{s}}_a(t)-\bm{u}(\bm{s}_a(t),t)\vert^2\Big\rangle\sim (u^\prime)^2\Big(L/\eta\Big)^{-2/3},\label{ChenK41} 
\end{align}
where $\bm{u}(\bm{s}_a(t),t)$ is the fluid velocity at $\bm{s}_a(t)$, ${u^\prime\equiv\sqrt{\langle\bm{u\cdot u}\rangle/3}}$ and $L$ is the integral lengthscale of the flow.  In the limit we are considering, namely ${Re_\lambda\to\infty}$, (\ref{ChenK41}) suggests that ${\dot{\bm{s}}_a(t)=\bm{u}(\bm{s}_a(t),t)}$, i.e. stagnation points are swept by the local fluid velocity. In \cite{coleman09} they use DNS to consider the joint PDF of $\dot{\bm{s}}_a(t)$ and $\bm{u}(\bm{s}_a(t),t)$ and do find a strong correlation, even at the modest values of Reynolds numbers in the study, ${Re_\lambda<200}$.  For ${St\ll1}$, ${\bm{v}^p(t)\approx\bm{u}(\bm{x}^p(t),t)-St\tau_\eta\bm{a}(\bm{x}^p(t),t)}$ where $\bm{v}^p(t)$ is the particle velocity and $\bm{u}(\bm{x}^p(t),t)$, $\bm{a}(\bm{x}^p(t),t)$ are the fluid velocity and acceleration at the particle position, respectively.  According to this expression, when $\bm{x}^p(t)=\bm{s}_a(t)$ the co-located particle moves with the fluid velocity $\bm{u}(\bm{x}^p(t),t)$.  This is statistically the same velocity with which the $\bm{a}=\bm{0}$ points move, and therefore it is argued that the particle sticks to $\bm{s}_a(t)$ and is swept along by $\bm{u}$. Although the above explanation for the stick part of the mechanism is technically valid only for $St\ll1$, in \cite{coleman09} they present results from DNS which, they argue, show that even for $St\sim1$, particles at acceleration stagnation points move, statistically, with the same velocity as the local fluid.

The conceptual framework of the sweep-stick mechanism is interesting and since particles do cluster near $\bm{a}=\bm{0}$ points, it provides a reasonable argument for inertial particle clustering. However, there is a confounding conceptual problem that occurs when applying the sweep-stick mechanism to stochastic flows such as kinematic simulations (KS). In KS, the acceleration stagnation points are uniformly distributed, yet the inertial particles still cluster. Chen \emph{et al.}~\cite{chen06} argued that clustering in this instance is due to the repelling action of the \emph{velocity} stagnation points (taken in the stationary frame of reference), which are clustered in KS.
%However, we would argue that, rather than concluding that the clustering mechanisms operating in DNS and KS are different, a more convincing conclusion would be that the sweep-stick mechanism is, in fact, not the underlying cause of the clustering in the inertial range.

However, the argument we presented in \S\ref{ACMIR} explains clustering in both KS and DNS. In particular, our argument states that the cause of the particle clustering lies in the nature of the interaction of the inertial particles with the fields $\widetilde{\bm{\mathcal{S}}}$ and $\widetilde{\bm{\mathcal{R}}}$.  This applies to both DNS and KS since it does not depend upon the dynamics of the underlying system governing $\widetilde{\bm{\mathcal{S}}}$ and $\widetilde{\bm{\mathcal{R}}}$. It is possible that the sweep-stick mechanism provides a valid explanation for clustering in DNS, but not KS, because of a relationship that exists between $\widetilde{\bm{\mathcal{S}}}$, $\widetilde{\bm{\mathcal{R}}}$ and $\bm{s}_a(t)$ that is specific to Navier-Stokes turbulence.  For example, in the limit $St\ll1$ \[\bm{\nabla_x\cdot}\bm{v}(\bm{x}^p(t),t)\approx-St\tau_\eta\Big(\mathcal{S}^2(\bm{x}^p(t),t)-\mathcal{R}^2(\bm{x}^p(t),t)\Big),\]which applies to any fluid velocity field that has spatial structure.  However, in Navier-Stokes turbulence\[\mathcal{S}^2(\bm{x}^p(t),t)-\mathcal{R}^2(\bm{x}^p(t),t)=-\bm{\nabla_x}^2 p^f(\bm{x}^p(t),t),\]such that in DNS one may speak of the behavior of ${\bm{\nabla_x\cdot}\bm{v}(\bm{x}^p(t),t)}$ in terms of either the particles interaction with $\bm{\mathcal{S}}$ and $\bm{\mathcal{R}}$, \emph{or equivalently} in terms of their interaction with the fluid pressure field $p^f$.  Yet, as the intrinsic clustering dynamics are due to the particle's interaction with $\bm{\mathcal{S}}$ and $\bm{\mathcal{R}}$, it is best to express ${\bm{\nabla_x\cdot}\bm{v}(\bm{x}^p(t),t)}$ in terms of those variables, since the result would be applicable to all flows.

It may well be the case that in an analogous way, a relationship exists in Navier-Stokes turbulence between $\widetilde{\bm{\mathcal{S}}}$,$\widetilde{\bm{\mathcal{R}}}$ and $\bm{s}_a(t)$.  A consequence of this could be that the explanations of inertial particle clustering in terms of either the clustering of $\bm{s}_a(t)$ points (as in the sweep-stick mechanism) or in terms of the particles preferential sampling of $\widetilde{\bm{\mathcal{S}}}$ over $\widetilde{\bm{\mathcal{R}}}$ (as in our explanation) are equivalent. To consider this possibility we will analyze the sweep-stick mechanism to see if it provides a relationship between $\bm{s}_a(t)$ and $\bm{x}^p(t)$.  We will then derive a relationship between $\widetilde{\bm{\mathcal{S}}}$, $\widetilde{\bm{\mathcal{R}}}$ and $\bm{s}_a(t)$ in Navier-Stokes turbulence and demonstrate that $\bm{s}_a(t)$ points cluster in regions where $\widetilde{\mathcal{A}}-\widetilde{\mathcal{B}}>0$, i.e., precisely the regions where the particles are predicted to cluster by the analysis in~\S\ref{ACMIR}.

\subsection{Generalization of the ``stick'' mechanism}

%Our principal concern with the sweep-stick mechanism is whether the ``stick'' part of the mechanism is really valid when ${St\gtrsim\mathcal{O}(1)}$.
The stick mechanism was formulated by appealing to the ${St\ll1}$ expression ${\bm{v}^p(t)=\bm{u}(\bm{x}^p(t),t)-St\tau_\eta\bm{a}(\bm{x}^p(t),t)}$; however, this expression is not valid for ${St\gtrsim\mathcal{O}(1)}$. In~\cite{coleman09}, they use DNS results to show that ${\bm{v}^p(t)=\bm{u}(\bm{x}^p(t),t)}$ when ${\bm{x}^p(t)=\bm{s}_a(t)}$.  Specifically, in \cite{coleman09} they show that ${\langle\bm{v}^p(t)-\bm{u}(\bm{x}^p(t),t)\rangle_{\bm{a}}=\bm{0}}$, when ${\bm{a}=\bm{0}}$, where ${\langle\cdot\rangle_{\bm{a}}}$ denotes an ensemble average conditioned on ${\bm{a}(\bm{x}^p(t),t)=\bm{a}}$. On this basis, they conclude that the stick mechanism is valid even for ${St\gtrsim\mathcal{O}(1)}$.  However, this result does not validate the stick mechanism for all Stokes numbers, nor does it explain the relationship between $\bm{v}^p(t)$ and $\bm{u}(\bm{x}^p(t),t)$ at ${\bm{a}=\bm{0}}$ points.  

Using the equation of motion we have
\begin{align}
-St\tau_\eta\Big\langle\dot{\bm{v}}^p(t)\Big\rangle_{\bm{a}}=\Big\langle\bm{v}^p(t)-\bm{u}(\bm{x}^p(t),t)\Big\rangle_{\bm{a}},	
\end{align}
and for the system of interest ${\langle\dot{\bm{v}}^p(t)\rangle=\bm{0}}$.  Deviations of ${\langle\dot{\bm{v}}^p(t)\rangle_{\bm{a}}}$ from $\bm{0}$ arise because of correlations between $\dot{\bm{v}}^p(t)$ and $\bm{a}(\bm{x}^p(t),t)$.  In the regime ${St\gtrsim\mathcal{O}(1)}$ where $\dot{\bm{v}}^p(t)$ is not uniquely defined by $\bm{a}(\bm{x}^p(t),t)$, $\bm{a}(\bm{x}^p(t),t)$ makes no contribution to $\dot{\bm{v}}^p(t)$ when ${\bm{x}^p(t)=\bm{s}_a(t)}$, meaning that at these points $\bm{a}(\bm{x}^p(t),t)$ and $\dot{\bm{v}}^p(t)$ are independent.  From this it follows that 
\begin{align}
\begin{split}
-St\tau_\eta\Big\langle\dot{\bm{v}}^p(t)\Big\rangle_{\bm{a}=\bm{0}}&\equiv-\frac{St\tau_\eta}{\varrho(\bm{0},t)}\Big\langle\dot{\bm{v}}^p(t)\delta(\bm{a}(\bm{x}^p(t),t)-\bm{0})\Big\rangle\\
&=-St\tau_\eta\Big\langle\dot{\bm{v}}^p(t)\Big\rangle\\
&=\bm{0}=\Big\langle\bm{v}^p(t)-\bm{u}(\bm{x}^p(t),t)\Big\rangle,
\end{split}
\end{align}
where ${\varrho(\bm{0},t)\equiv\langle\delta(\bm{a}(\bm{x}^p(t),t)-\bm{0})\rangle}$.
That the mean particle and fluid velocities at $\bm{s}_a(t)$ points are equal does not validate the stick mechanism since two variables with equal expectations may be statistically independent of one another.  Furthermore,\[\lim_{St\to\infty}St\tau_\eta\Big\langle\dot{\bm{v}}^p(t)\Big\rangle_{\bm{a}=\bm{0}}=\bm{0},\]which, if ${\langle\bm{v}^p(t)-\bm{u}(\bm{x}^p(t),t)\rangle_{\bm{a}=\bm{0}}=\bm{0}}$ were sufficient to demonstrate the stick mechanism, would imply that ${St\to\infty}$ particles should cluster through the action of the sweep-stick mechanism, which is clearly invalid \footnote[2]{Here we are taking the limit $St\to\infty$ with $\tau_p\to\infty$ and $\tau_\eta$ finite, in which limit the particles do not cluster.  It should be noted however that if we instead take the limit $St\to\infty$ with $\tau_p$ finite but $\tau_\eta\to0$, as in a white-in-time flow, then the particles can still cluster \cite{gustavsson11b}.  Constructing the limit in this second way is however not relevant to our discussion of the sweep-stick mechanism which is concerned with real turbulence.}.

In order to demonstrate that the stick mechanism is valid for ${St\gtrsim\mathcal{O}(1)}$ one must consider a statistic such as
\begin{align}
\mathcal{Q}\equiv\Big\langle\vert\bm{v}^p(t)-\bm{u}(\bm{x}^p(t),t)\vert^2\Big\rangle_{\bm{a}}=(St\tau_\eta)^2\Big\langle\vert\dot{\bm{v}}^p(t)\vert^2\Big\rangle_{\bm{a}},\label{Q}
\end{align}
which is only zero at ${\bm{a}=\bm{0}}$ if ${\bm{v}^p(t)=\bm{u}(\bm{x}^p(t),t)}$.  
In the regime ${St\ll1}$, ${\mathcal{Q}=(St\tau_\eta)^2\vert\bm{a}\vert^2}$, which is consistent with the stick mechanism.  However, as explained earlier, in the regime ${St\gtrsim\mathcal{O}(1)}$, $\bm{a}(\bm{x}^p(t),t)$ and $\dot{\bm{v}}^p(t)$ are independent when ${\bm{x}^p(t)=\bm{s}_a(t)}$, and since ${\langle\vert\dot{\bm{v}}^p(t)\vert^2\rangle\neq\bm{0}}$ then ${\mathcal{Q}(\bm{a}=\bm{0})\neq{0}}$.  Nevertheless, in order for the stick mechanism to be valid one does not necessarily require that ${\mathcal{Q}(\bm{a}=\bm{0})=0}$ precisely but rather that ${\mathcal{Q}(\bm{a}=\bm{0})}$ is in some sense small.  For example, the sweep part of the mechanism suggests that the velocity with which the $\bm{s}_a(t)$ points are swept is related to $u^\prime$.  In this case, if ${\mathcal{Q}(\bm{a}=\bm{0})\ll u^\prime u^\prime}$, then even though the particles do not precisely stick to the stagnation points, they remain close enough to follow them in a significant way.

%\vspace{-1mm}
\begin{figure}[t]
\centering
{\begin{overpic}
[trim = 20mm 60mm 20mm 60mm,scale=0.45,clip,tics=20]{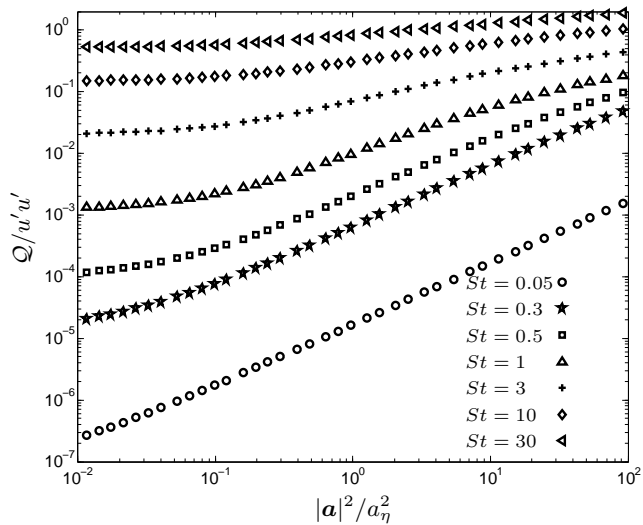}
\put(-13,95){\rotatebox{90}{$\mathcal{Q}/u^\prime u^\prime$}}
\put(102,0){$\vert\bm{a}\vert^2/a_\eta^2$}
\put(159,86){\scriptsize\text{$St=0.05$}}
\put(159,76){\scriptsize\text{$St=0.3$}}
\put(159,66){\scriptsize\text{$St=0.5$}}
\put(159,56){\scriptsize\text{$St=1$}}
\put(159,46){\scriptsize\text{$St=3$}}
\put(159,36){\scriptsize\text{$St=10$}}
\put(159,26){\scriptsize\text{$St=30$}}
\end{overpic}}
\caption{DNS data for $\mathcal{Q}$ at various $St$, plotted as a function of $\vert\bm{a}\vert^2/a_\eta^2$, where $a_\eta$ is the Kolmogorov acceleration.}
\label{Q_DNS}
\end{figure}
%\FloatBarrier

In Figure~\ref{Q_DNS} we show results for $\mathcal{Q}$ computed from DNS at $Re_\lambda=597$.  Details on the DNS used throughout this paper can be found in \cite{ireland14}.  As expected, the results show that ${\mathcal{Q}=(St\tau_\eta)^2\vert\bm{a}\vert^2}$ for ${St\ll1}$, implying ${\mathcal{Q}(\bm{a}\to\bm{0})\to0}$, consistent with the stick mechanism.  For ${St=\mathcal{O}(1)}$, while ${\mathcal{Q}(\bm{a}\to\bm{0})\not\to0}$, ${\mathcal{Q}(\bm{a}\to\bm{0})\ll u^\prime u^\prime}$, implying that although the particles do not precisely stick to $\bm{s}_a(t)$ points, they remain close enough to follow them in a significant way.  For ${St=\mathcal{O}(10)}$, ${\mathcal{Q}(\bm{a}\to\bm{0})}$ remains quite small relative to $u^\prime u^\prime$.  However, for ${St=\mathcal{O}(10)}$ the variation of ${\mathcal{Q}}$ with $\bm{a}$ for ${\vert\bm{a}\vert^2/a_\eta^2\leq\mathcal{O}(1)}$ is weak.  This implies that although ${\mathcal{Q}(\bm{a}\to\bm{0})}$ is still smaller than $u^\prime u^\prime$ at ${St=\mathcal{O}(10)}$, the significance of $\bm{s}_a(t)$ points for the particle motion becomes small.  This follows from noting that if ${\mathcal{Q}(\bm{a})}$ were constant for a given $St$, then it would imply that the particle motion is entirely uncorrelated with $\bm{a}(\bm{x}^p(t),t)$.  Nevertheless, our DNS data shows that ${St=\mathcal{O}(10)}$ particles cluster, and in fact cluster more strongly in the inertial range than ${St=\mathcal{O}(1)}$ particles (see \cite{ireland14}), indicating the breakdown of the sweep-stick mechanism as the explanation for clustering when $St=\mathcal{O}(10)$.  In our DNS at ${Re_\lambda=597}$, ${St\lesssim\mathcal{O}(1)\implies St_r\ll1}$, and ${St\gtrsim\mathcal{O}(10)\implies St_r\gtrsim\mathcal{O}(1)}$ for $r$ in the inertial range.  The conclusion to be drawn from Figure~\ref{Q_DNS} is then that the sweep-stick mechanism provides a valid explanation for clustering in the inertial range of Navier-Stokes turbulence when ${St_r\ll1}$, but it does not apply when ${St_r\gtrsim\mathcal{O}(1)}$.  This is not surprising since the sweep-stick mechanism is essentially a local mechanism.

Next we consider the relationship between the sweep-stick mechanism and the mechanism presented in \S\ref{ACMIR}, in the limit $St_r\ll 1$. If they are related, we should be able to demonstrate that in Navier-Stokes turbulence $\bm{s}_a(t)$ points cluster in regions where ${\widetilde{\mathcal{A}}-\widetilde{\mathcal{B}}>0}$, which are the same regions where the particles are predicted to cluster by our analysis in \S\ref{ACMIR}.

\subsection{Where do $\bm{s}_a(t)$ points cluster?}

We begin by defining the PDF $\mathcal{P}(\bm{r},\bm{w},t)\equiv\Big\langle\delta(\Delta\bm{s}_a(t)-\bm{r})\delta(\Delta\dot{\bm{s}}_a(t)-\bm{w})\Big\rangle$, whose exact evolution equation is
\begin{align}
\partial_t \mathcal{P}=-\bm{\nabla_r\cdot}\mathcal{P}\bm{w}-\bm{\nabla_w\cdot}\mathcal{P}\Big\langle\Delta\ddot{\bm{s}}_a(t)\Big\rangle_{\bm{r},\bm{w}},\label{PDFsa}
\end{align}
where $\Delta\bm{s}_a(t)$, $\Delta\dot{\bm{s}}_a(t)$ and $\Delta\ddot{\bm{s}}_a(t)$ are the relative separation, relative velocity and relative acceleration vectors between the location of two stagnation points, respectively.  From (\ref{PDFsa}) we can derive the exact equation governing the statistically stationary distribution of $\Delta\bm{s}_a(t)$, namely  the equation governing ${\varrho(\bm{r})\equiv\int_{\bm{w}}\mathcal{P}(\bm{r},\bm{w})\,d\bm{w}}$
\begin{align}
\begin{split}
\bm{0}=&-\Big\langle\Delta\dot{\bm{s}}_a(t)\Delta\dot{\bm{s}}_a(t)\Big\rangle_{\bm{r}}\bm{\cdot\nabla_r}\varrho+\varrho\Big\langle\Delta\ddot{\bm{s}}_a(t)\Big\rangle_{\bm{r}}\\
&-\varrho\bm{\nabla_r\cdot}\Big\langle\Delta\dot{\bm{s}}_a(t)\Delta\dot{\bm{s}}_a(t)\Big\rangle_{\bm{r}}.\label{RDFsa}
\end{split}
\end{align}
In order to proceed we need to know something about the dynamics of the turbulence at the $\bm{s}_a(t)$ points.  According to the sweep mechanism, which is based upon a K41 description of the turbulence dynamics, ${\Delta\ddot{\bm{s}}_a(t)\approx\bm{0}}$ and ${\Delta\dot{\bm{s}}_a(t)\approx\Delta\bm{u}(\Delta\bm{s}_a(t),t)}$ so that (\ref{RDFsa}) becomes
\begin{align}
\begin{split}
\bm{0}=&-\Big\langle\Delta\bm{u}(\Delta\bm{s}_a(t),t)\Delta\bm{u}(\Delta\bm{s}_a(t),t)\Big\rangle_{\bm{r}}\bm{\cdot\nabla_r}\varrho\\
&-\varrho\bm{\nabla_r\cdot}\Big\langle\Delta\bm{u}(\Delta\bm{s}_a(t),t)\Delta\bm{u}(\Delta\bm{s}_a(t),t)\Big\rangle_{\bm{r}},
\end{split}
\label{RDFsa2}
\end{align}
where $\Delta\bm{u}(\Delta\bm{s}_a(t),t)$ is the vector difference between the fluid velocity at the positions of the two stagnation points.  The drift flux in (\ref{RDFsa2}) has precisely the same form as the term appearing in the drift velocity describing inertial particle clustering in the limit ${St_r\ll1}$, except that now the fluid velocity increments are measured at $\Delta\bm{s}_a(t)$ instead of $\bm{r}^p(t)$ (see \S\ref{ACMIR}).  Consequently, we may use the same coarse-graining analysis to re-express the drift flux in (\ref{RDFsa2}) in terms of $\widetilde{\bm{\mathcal{S}}}$ and $\widetilde{\bm{\mathcal{R}}}$.  Doing this, we arrive at the following result for ${\eta\ll r\ll L}$
\begin{align}
\bm{\nabla_r\cdot}\Big\langle\Delta\bm{u}(\Delta\bm{s}_a(t),t)\Delta\bm{u}(\Delta\bm{s}_a(t),t)\Big\rangle_{\bm{r}}\approx\frac{7}{45}\bm{r}(\widetilde{\mathcal{A}}-\widetilde{\mathcal{B}}),\label{Adrift2}	
\end{align}
where now the coarse-grained invariants $\widetilde{\mathcal{A}}$ and $\widetilde{\mathcal{B}}$ are based on $\bm{\mathcal{S}}(\bm{s}_a(t),t)$ and $\bm{\mathcal{R}}(\bm{s}_a(t),t)$ (i.e. strain-rate and rotation-rate measured at $\bm{s}_a(t)$ instead of $\bm{x}^p(t)$).  Just as (\ref{eq:drift}) was derived under the assumption that at ${St_r\ll1}$ the particle clustering is weak, (\ref{Adrift2}) assumes that the clustering of $\bm{s}_a(t)$ points is weak in the inertial range, as is indicated by the DNS results in \cite{chen06}.

The result in (\ref{Adrift2}), when inserted into (\ref{RDFsa2}), demonstrates that $\bm{s}_a(t)$ points drift into and cluster in regions where $\widetilde{\mathcal{A}}-\widetilde{\mathcal{B}}>0$.  The mechanism by which they drift into these regions is connected to the turbulence dynamics, and in particular the nonlinear sweeping effect which generates ${\Delta\dot{\bm{s}}_a(t)\approx\Delta\bm{u}(\Delta\bm{s}_a(t),t)}$.  

\vspace{-1mm}
\begin{figure}[t]
\centering
{\begin{overpic}
[trim = 20mm 70mm 20mm 70mm,scale=0.45,clip,tics=20]{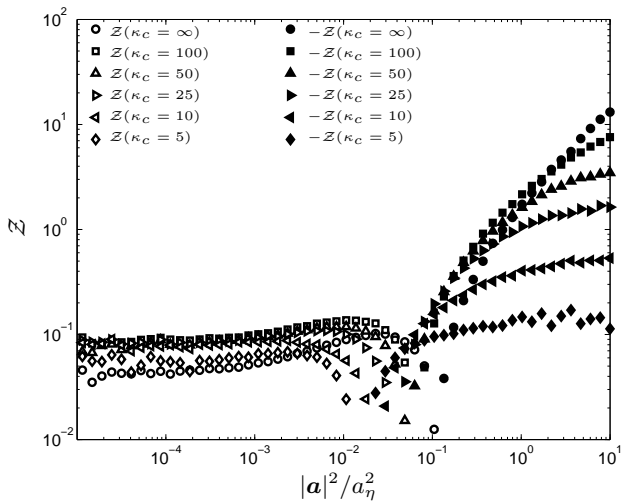}
\put(-10,90){\rotatebox{90}{$\mathcal{Z}$}}
\put(100,-7){$\vert\bm{a}\vert^2/a_\eta^2$}
\put(28,166){\tiny\text{$\mathcal{Z}(\kappa_c=\infty)$}}
\put(103,166){\tiny\text{$-\mathcal{Z}(\kappa_c=\infty)$}}
\put(28,158){\tiny\text{$\mathcal{Z}(\kappa_c=100)$}}
\put(103,158){\tiny\text{$-\mathcal{Z}(\kappa_c=100)$}}
\put(28,150){\tiny\text{$\mathcal{Z}(\kappa_c=50)$}}
\put(103,150){\tiny\text{$-\mathcal{Z}(\kappa_c=50)$}}
\put(28,142){\tiny\text{$\mathcal{Z}(\kappa_c=25)$}}
\put(103,142){\tiny\text{$-\mathcal{Z}(\kappa_c=25)$}}
\put(28,134){\tiny\text{$\mathcal{Z}(\kappa_c=10)$}}
\put(103,134){\tiny\text{$-\mathcal{Z}(\kappa_c=10)$}}
\put(28,126){\tiny\text{$\mathcal{Z}(\kappa_c=5)$}}
\put(103,126){\tiny\text{$-\mathcal{Z}(\kappa_c=5)$}}
\end{overpic}}
\caption{DNS data for $\mathcal{Z}$ at various cut-off wavenumbers $\kappa_c$, plotted as a function of $\vert\bm{a}\vert^2/a_\eta^2$.}
\label{S2_R2_con_a_plot}
\end{figure}
%\FloatBarrier

In order to confirm this prediction that $\bm{s}_a(t)$ points are associated with regions where ${\widetilde{\mathcal{A}}-\widetilde{\mathcal{B}}>0}$ we computed the quantity ${\mathcal{Z}\equiv\langle\widetilde{\bm{\mathcal{S}}}\bm{:}\widetilde{\bm{\mathcal{S}}}-\widetilde{\bm{\mathcal{R}}}\bm{:}\widetilde{\bm{\mathcal{R}}}\rangle_{\bm{a}}}$ using DNS.  The coarse-graining was performed using a sharp spectral cut-off at wavenumber $\kappa_c$.  The results in Figure~\ref{S2_R2_con_a_plot} confirm the prediction in (\ref{Adrift2}) since they show that regions where the fluid acceleration is low (${\bm{a}\to\bm{0}}$) are associated with regions where the coarse-grained strain exceeds the coarse-grained rotation (${\mathcal{Z}>0}$).

%We have shown then that $\bm{s}_a(t)$ and $\bm{x}^p(t)$ both cluster in Navier-Stokes turbulence in regions where ${\widetilde{\mathcal{A}}-\widetilde{\mathcal{B}}>0}$.  The sweep-stick mechanism provides an explanation for why the clustering of $\bm{s}_a(t)$ and $\bm{x}^p(t)$ are correlated in the inertial range if ${St_r\ll1}$.  However, in purely kinematic flow fields where nonlinear sweeping effects are absent, the particles still cluster in regions where ${\widetilde{\mathcal{A}}-\widetilde{\mathcal{B}}>0}$ but $\bm{s}_a(t)$ do not, as observed in the KS results in \cite{chen06}.  Consequently, it is the particles interaction with $\widetilde{\bm{\mathcal{S}}}$ and $\widetilde{\bm{\mathcal{R}}}$, not $\bm{s}_a(t)$, that is the real cause of their clustering.  The sweep-stick mechanism only provides an explanation in real turbulence because of the particular relationship that exists between $\bm{s}_a(t)$ and $\widetilde{\bm{\mathcal{S}}}$,$\widetilde{\bm{\mathcal{R}}}$ in Navier-Stokes dynamics.

In closing this section we note that the prediction in \S\ref{ACMIR} that the inertial particles cluster in regions where ${\widetilde{\mathcal{A}}-\widetilde{\mathcal{B}}>0}$ is only guaranteed for ${St_r\ll1}$, where the drift velocity is given by (\ref{eq:drift}).  When ${St_r\gtrsim\mathcal{O}(1)}$ the non-local clustering mechanism contributes, and indeed dominates the centrifuge mechanism in the inertial range in the limit ${Re_\lambda\to\infty}$.  When the non-local clustering mechanism dominates it is much more complicated to predict theoretically where the particles will cluster in the flow.  However, recent work has shown that the non-local clustering mechanism in the dissipation range causes the particles to accumulate in the same high-strain, low-rotation regions of the turbulence as the local mechanism~\cite{bragg14d}. The analysis can be ported over to the inertial range, but now using the coarse-grained fluid velocity gradient field, to show that in the limit ${Re_\lambda\to\infty}$ and when ${St_r\gtrsim\mathcal{O}(1)}$, the particles still cluster in regions where ${\widetilde{\mathcal{A}}-\widetilde{\mathcal{B}}>0}$.  

\section{Predicting the RDF in the inertial range}\label{RDF_pred}

In \S\ref{ACMIR} we analyzed the exact equation governing $g(r)$ in order to consider the mechanism generating clustering when ${\eta\ll r\ll L}$.  In this section we use a closed model equation for $g(r)$ in order to predict the functional form of $g(r)$ in the inertial range, in the limit ${St_r\ll1}$.

For isotropic turbulence, (\ref{RDFeq}) may be re-written as
\begin{align}
\begin{split}
0=&g\langle\Delta u_\parallel(r^p(t),t)\rangle_r-St\tau_\eta S^p_{2\parallel}\nabla_r g\\
&-St\tau_\eta g\Big(\nabla_r S^p_{2\parallel}+2r^{-1}[S^p_{2\parallel}-S^p_{2\perp}]\Big),\label{RDFeqISO}
\end{split}
\end{align}
where the subscripts $\parallel$ and $\perp$ denote the longitudinal and transverse projections of the tensors and ${r^p(t)=\vert\bm{r}^p(t)\vert}$.  In \cite{zaichik07} the term ${\langle\Delta u_\parallel(r^p(t),t)\rangle_r}$ is closed by approximating $\Delta\bm{u}(\bm{r},t)$ as a spatio-temporally correlated Gaussian field and by using the Furutsu-Novikov closure method.  The result they obtained was
\begin{align}
\langle\Delta u_\parallel(r^p(t),t)\rangle_r\approx-\frac{1}{g}St\tau_\eta\lambda_\parallel\nabla_r g,\label{ZTclosure}	
\end{align}
and for ${St_r\ll1}$, ${\eta\ll r\ll L}$
\begin{align}
{\lambda_\parallel=(St\tau_\eta)^{-1}\gamma C_2\langle\epsilon\rangle^{1/3}r^{4/3}},\label{lambda_C}
\end{align}
where ${C_2=2.1}$~\cite{sreeni95}, ${\gamma=\tau_\mathcal{S}(15 C_2\tau_\eta^2)^{-1/2}}$ \cite{zaichik03} and $\tau_\mathcal{S}$ is the Lagrangian timescale of $\bm{\mathcal{S}}$.  In our DNS $\tau_\mathcal{S}=2.02\tau_\eta$. 

It is well known that in turbulence $\Delta u_\parallel(r,t)$ can be strongly non-Gaussian, which calls into question the closure result in (\ref{ZTclosure}).  However, results in \cite{bragg14b} indicate that even for ${r\ll\eta}$, neglecting the non-Gaussian features of $\Delta u_\parallel(r,t)$ in the closure of $\langle\Delta u_\parallel(r^p(t),t)\rangle_r$ has a negligible effect on $g(r)$.  This is likely a consequence of the fact that $g(r)$ is a low-order moment of the particle phase-space dynamics and therefore that it is only weakly affected by the strongly non-Gaussian features of $\Delta u_\parallel(r,t)$, which predominantly manifest themselves in the tails of the distribution.  Therefore, for the present purposes of using the closure in (\ref{ZTclosure}) for ${\eta\ll r\ll L}$, the neglect of the non-Gaussianity of $\Delta u_\parallel(r,t)$ in the closure should be even less important since the non-Gaussianity of $\Delta u_\parallel(r,t)$ is weaker in the inertial range than in the dissipation range \cite{ishihara09}.

\begin{figure}[t]
\centering
{\begin{overpic}
[trim = 20mm 70mm 20mm 70mm,scale=0.5,clip,tics=20]{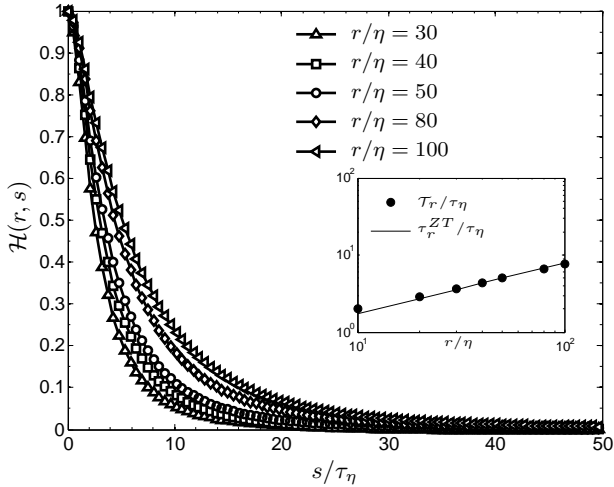}
\put(5,90){\rotatebox{90}{$\mathcal{H}(r,s)$}}
\put(120,5){$s/\tau_\eta$}
\put(135,171){\footnotesize\text{$r/\eta=30$}}
\put(135,160){\footnotesize\text{$r/\eta=40$}}
\put(135,149){\footnotesize\text{$r/\eta=50$}}
\put(135,138){\footnotesize\text{$r/\eta=80$}}
\put(135,127){\footnotesize\text{$r/\eta=100$}}
\put(170,55){\tiny\text{$r/\eta$}}
\put(160,108){\tiny\text{$\mathcal{T}_r/\tau_\eta$}}
\put(160,98){\tiny\text{$\tau^{ZT}_r/\tau_\eta$}}
\end{overpic}}
\caption{DNS data for $\mathcal{H}$ for various $r$ as a function of $s$.  The inset shows a comparison of the timescale ${\mathcal{T}_r\equiv\int_0^\infty\mathcal{H}ds}$ with the ZT prediction ${\tau^{ZT}_r=\gamma\langle\epsilon\rangle^{-1/3}r^{2/3}}$.}
\label{Auto_IR}
\end{figure}

In deriving the closed expression for ${\lambda_\parallel}$ given in (\ref{lambda_C}), ZT approximated the Lagrangian autocovariances of $\Delta\bm{u}(\bm{r}^p(t),t)$ as having an exponential decay in time with the timescale given by ${\tau^{ZT}_r=\gamma\langle\epsilon\rangle^{-1/3}r^{2/3}}$. However, this appears to be in conflict with the behavior one would expect based on K41 arguments, namely
\begin{align}
\Big\langle\Delta\bm{u}(\bm{r}^p(0),0)\bm{\cdot}\Delta\bm{u}(\bm{r}^p(s),s)\Big\rangle_{\bm{r}}\propto \langle\epsilon\rangle s,\label{uuK41}	
\end{align}
for ${St=0}$, according to which the autocovariances should grow indefinitely in the inertial range as ${Re_\lambda\to\infty}$.  However, it is known that applications of K41 scaling arguments to Lagrangian statistics can be in significant error, even for low order moments \cite{falkovich12}.  In Fig.~\ref{Auto_IR} we show results computed from our DNS for \[\mathcal{H}(r,s)\equiv\frac{\langle\Delta\bm{u}(\bm{r}^p(0),0)\bm{\cdot}\Delta\bm{u}(\bm{r}^p(s),s)\rangle_{r}}{\langle\Delta\bm{u}(\bm{r}^p(0),0)\bm{\cdot}\Delta\bm{u}(\bm{r}^p(0),0)\rangle_{r}},\] for $St=0$ particles at ${\eta\ll r\ll L}$.  

\vspace{-1mm}
%\FloatBarrier

The results show that $\mathcal{H}$ is in fact a decaying function of $s$ at ${\eta\ll r\ll L}$ and therefore demonstrate that (\ref{uuK41}) is fundamentally incorrect.  We expect that the failure of the prediction in (\ref{uuK41}) is due to the fact that such a simple scaling argument does not capture the effect of the spatio-temporal decorrelation of the velocity field along the pair trajectory, and only accounts for the fact that as the pair separates, the two-point, one-time fluid velocity increments increase along the pair trajectory.  In the inset of Fig.~\ref{Auto_IR} we compare ${\mathcal{T}_r\equiv\int_0^\infty\mathcal{H}ds}$ with the ZT prediction ${\tau^{ZT}_r=\gamma\langle\epsilon\rangle^{-1/3}r^{2/3}}$ which is used in their closure for $\lambda_\parallel$.  The results show a remarkable agreement between $\tau^{ZT}_r$ and $\mathcal{T}_r$ and confirm the validity of the closure approximation made in the ZT for $\lambda_\parallel$ when ${\eta\ll r\ll L}$.  

If we now substitute (\ref{ZTclosure}) into (\ref{RDFeqISO}) and also use the result in (\ref{eq:drift}) for the isotropic form of ${St\tau_{\eta}\boldsymbol{\nabla_r\cdot}\boldsymbol{S}^{p}_{2}}$ for ${St_r\ll1}$ and ${\eta\ll r\ll L}$, we obtain the solution
\begin{align}
g(r)=\exp\Bigg[-\frac{7 St\tau_\eta}{45\gamma C_2\langle\epsilon\rangle^{1/3}}\int\limits_{0}^{r} \mathfrak{r}^{-1/3}(\widetilde{\mathcal{A}}-\widetilde{\mathcal{B}})d\mathfrak{r}\Bigg].\label{CG_RDF}
\end{align}
The expression in (\ref{CG_RDF}) requires knowledge of $\widetilde{\mathcal{A}}-\widetilde{\mathcal{B}}$, which is difficult to predict.  However, we can obtain an approximation for its $r$ dependence in the limit ${St_r\ll1}$, which allows us through (\ref{CG_RDF}) to determine the $r$ dependence of $g(r)$ over the range ${\eta St^{3/2} \ll r\ll L}$. In this limit, we introduce a perturbation expansion for ${\widetilde{\mathcal{A}}-\widetilde{\mathcal{B}}}$ in $St_r$
\begin{align}
\widetilde{\mathcal{A}}-\widetilde{\mathcal{B}}=[\widetilde{\mathcal{A}}-\widetilde{\mathcal{B}}]^{[0]}+St_r[\widetilde{\mathcal{A}}-\widetilde{\mathcal{B}}]^{[1]}+\mathcal{O}(St_r^2),	
\end{align}
where the superscript $[\cdot]$ denotes the order of the perturbation term. The zeroth-order term, ${[\widetilde{\mathcal{A}}-\widetilde{\mathcal{B}}]^{[0]}}$, which represents ${\widetilde{\mathcal{A}}-\widetilde{\mathcal{B}}}$ measured along fluid particle trajectories, is zero. Based on K41, we expect that to leading order in $St_r$, ${[\widetilde{\mathcal{A}}-\widetilde{\mathcal{B}}]^{[1]}\propto r^{-4/3}}$, and using this together with the definition for $St_r$, which can be re-expressed as ${St_r\equiv St(r/\eta)^{-2/3}}$, we obtain
\begin{align}
\begin{split}
\widetilde{\mathcal{A}}-\widetilde{\mathcal{B}}&=St(r/\eta)^{-2/3}[\widetilde{\mathcal{A}}-\widetilde{\mathcal{B}}]^{[1]}+\mathcal{O}(St_r^2)\\
&\propto r^{-2}.\label{AmBpred}
\end{split}
\end{align}
Substituting this into (\ref{CG_RDF}), we arrive at the following expression for $g(r)$ in the limit $St_r\ll1$
\begin{align}
g(r)=\exp[\mathcal{D}r^{-4/3}],\label{g_r_dep}
\end{align}
where $\mathcal{D}$ is an unknown positive coefficient that is independent of $r$, but dependent on Stokes number, satisfying ${\mathcal{D}(St=0)=0}$.    

Equation (\ref{g_r_dep}) implies that even for ${St_r\ll 1}$, clustering in the inertial range is not scale-invariant \cite{bec07,pan11}\footnote[3]{In \cite{bragg14c} we incorrectly argued that $g(r)$ is approximately a power law at ${\eta\ll r\ll L}$ when ${St_r\ll1}$ since we mistakingly assumed ${\lambda_\parallel\propto S^p_{2\parallel}}$ in this regime.}, in contrast to clustering in the dissipation range for ${St\ll1}$.  This may seem surprising given that we argued that the mechanism generating the clustering in the inertial range is completely analogous to the mechanism in the dissipation range (cf. \S\ref{ACMIR}). The difference in the form of the clustering does not arise from a difference in the mechanism generating the clustering.  Note also that according to our analysis the break in the scale-invariance of the particle clustering in the inertial range has nothing to do with the breakdown of the scale-invariance of $\Delta\bm{u}(\bm{r},t)$ in the inertial range \cite{falkovich09} since our analysis used K41 scaling.  The break in the scale-invariance of the clustering going from the dissipation to the inertial range is simply a consequence of the fact that $\tau_r$ is dependent on $r$ in the inertial range, but is independent of $r$ in the dissipation range.  The final steady state form of $g(r)$ depends upon the way the drift and diffusion processes depend upon $r$, and their relative scaling with $r$ is different in the dissipation and inertial ranges precisely because of the behavior of $\tau_r$.

\begin{figure}[t]
\centering
{\begin{overpic}
[trim = 20mm 70mm 20mm 70mm,scale=0.5,clip,tics=20]{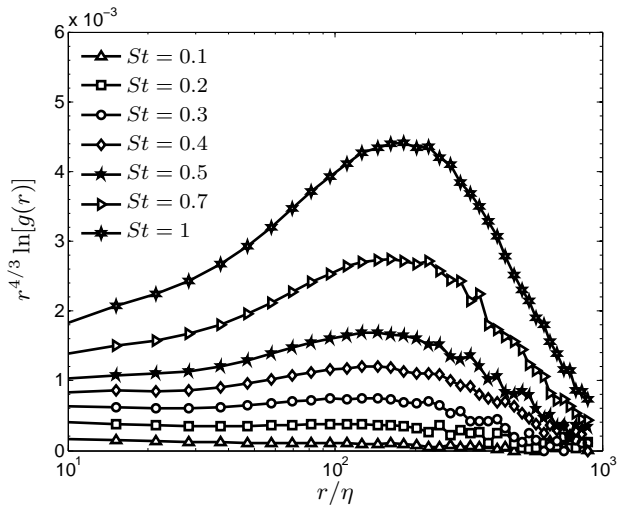}
\put(5,80){\rotatebox{90}{$r^{4/3}\ln[g(r)]$}}
\put(122,5){$r/\eta$}
\put(50,170){\footnotesize\text{$St=0.1$}}
\put(50,159){\footnotesize\text{$St=0.2$}}
\put(50,148){\footnotesize\text{$St=0.3$}}
\put(50,137){\footnotesize\text{$St=0.4$}}
\put(50,126){\footnotesize\text{$St=0.5$}}
\put(50,115){\footnotesize\text{$St=0.7$}}
\put(50,104){\footnotesize\text{$St=1$}}
\end{overpic}}
\caption{DNS data for $r^{4/3}\ln[g(r)]$ for various $St$ as a function of $r$.}
\label{IR_RDF_plot}
\end{figure}

In Figure~\ref{IR_RDF_plot}, we use DNS data to test the prediction in (\ref{g_r_dep}) by plotting ${r^{4/3}\ln[g(r)]}$. In these coordinates, (\ref{g_r_dep}) implies a horizontal line in the inertial range. The results show that the predicted form in (\ref{g_r_dep}) is quite accurate for ${St\lesssim0.3}$ and $10\eta\lesssim r\lesssim 200\eta$. Deviations from (\ref{g_r_dep}) for ${St>0.3}$, over the same range of separations, are due to the breakdown of the predicted scaling ${\widetilde{\mathcal{A}}-\widetilde{\mathcal{B}}\propto r^{-2}}$.  If we assume in general ${\widetilde{\mathcal{A}}-\widetilde{\mathcal{B}}\propto r^{-\alpha}}$ then $g(r)$ would take the form ${g(r)=\exp[\mathcal{D}r^{(2-3\alpha)/3}]}$.  Our data indicates that over the range of $r$ that we have access to in our DNS, $\alpha\leq2$, and this explains why the results in Figure~\ref{IR_RDF_plot} show that for ${St>0.3}$ and ${10\eta\lesssim r\lesssim 200\eta}$, ${\nabla_r(r^{4/3}\ln[g(r)])>0}$.  

%\vspace{-3mm}
%
%\FloatBarrier

The results in Figure~\ref{IR_RDF_plot} for ${200\eta\lesssim r\lesssim L}$ show that for all $St$, ${\nabla_r(r^{4/3}\ln[g(r)])<0}$. This deviation of $g(r)$ from the form predicted in (\ref{g_r_dep}) cannot be due to a breakdown of the validity of the perturbation analysis used to derive (\ref{g_r_dep}), as this approximation should improve as $r$ increases. 
The cause is actually the influence of the large scales. The DNS data shows that $\Delta\bm{u}(\bm{r},t)$ begins to depart from its inertial range scaling at ${r\approx 200\eta}$. Naturally this transition is eliminated in the limit ${Re_\lambda\to\infty}$.

\begin{figure}[t]
\centering
{\begin{overpic}
[trim = 20mm 78mm 20mm 78mm,scale=0.5,clip,tics=20]{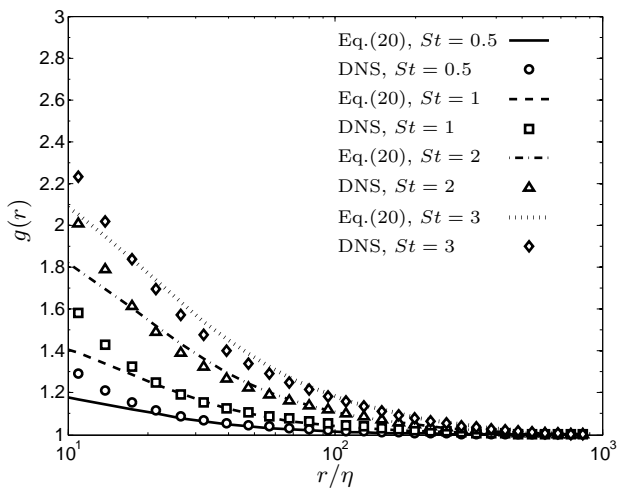}
\put(5,85){\rotatebox{90}{$g(r)$}}
\put(130,148){\scriptsize\text{DNS, $St=0.5$}}
\put(130,159){\scriptsize\text{Eq.(\ref{CG_RDF}), $St=0.5$}}
\put(130,126){\scriptsize\text{DNS, $St=1$}}
\put(130,137){\scriptsize\text{Eq.(\ref{CG_RDF}), $St=1$}}
\put(130,104){\scriptsize\text{DNS, $St=2$}}
\put(130,115){\scriptsize\text{Eq.(\ref{CG_RDF}), $St=2$}}
\put(130,81){\scriptsize\text{DNS, $St=3$}}
\put(130,92){\scriptsize\text{Eq.(\ref{CG_RDF}), $St=3$}}
\put(122,-6){$r/\eta$}
\end{overpic}}
\caption{Plot of DNS data and the predictions of (\ref{CG_RDF}) for $g(r)$.}
\label{CG_ZT_DNS}
\end{figure}

In order to test the quantitative accuracy of (\ref{CG_RDF}) we evaluate $\widetilde{\mathcal{A}}-\widetilde{\mathcal{B}}$ from the DNS using a sharp spectral cut-off at wavenumber ${\kappa_c=2\pi/r}$ for the coarse-graining. Figure~\ref{CG_ZT_DNS} compares $g(r)$ directly computed from the DNS with that obtained from (\ref{CG_RDF}) using DNS data for $\widetilde{\mathcal{A}}-\widetilde{\mathcal{B}}$. The results demonstrate the accuracy of (\ref{CG_RDF}) over the inertial range, in the limit ${St_r\ll1}$.  At this $Re_\lambda$ (597), $St>3$ particles do not satisfy the ${St_r\ll1}$ requirement in the inertial range.

%\vspace{-3.2cm}
%\FloatBarrier

Finally, we consider the behavior of $g(r)$ in the limit ${Re_\lambda\to\infty}$ as $r$ decreases. For ${St\lesssim\mathcal{O}(1)}$, $g(r)$ will transition from (\ref{CG_RDF}) to the scale-invariant form ${g(r)\propto r^{-\xi(St)}}$ at ${r\ll\eta}$, where ${\xi(St)\ge 0}$. For ${St\gg 1}$, $g(r)$ will deviate from (\ref{CG_RDF}) at ${\eta\ll r\sim St^{3/2}\eta\ll L}$.  At ${r\sim St^{3/2}\eta}$, ${St_r=\mathcal{O}(1)}$ at which point the path-history symmetry breaking effect dominates the clustering mechanism.  We cannot derive a prediction for the analytic form of $g(r)$ in this regime because the particle relative velocity structure function in this regime is not a simple power law.  As $r$ decreases further, the particles enter a ballistic regime, where ${g(r)\approx\text{constant}}$~\cite{pan11,bragg14c}. All of these trends can be seen in \cite{ireland14}. The theoretical question of the existence of a transition to ${g(r)\approx\text{constant}}$ for ${St\lesssim\mathcal{O}(1)}$ at ${r\lll\eta}$ remains an open question~\cite{bragg14c}.
\FloatBarrier

\section{Conclusions}

In this paper, we have considered the mechanism for the clustering of inertial particles in the inertial range of isotropic turbulence.  By analyzing the exact equation governing the RDF, we have demonstrated that the clustering mechanisms in the inertial range are completely analogous to the mechanisms in the dissipation range. For any separation $r$ which is less than the integral lengthscale of the flow, the clustering mechanism for ${St_r\ll1}$ is related to the preferential sampling of the coarse-grained fluid velocity gradient tensor at scale $\sim r$, which is associated with centrifuging out of eddies at that scale.  When ${St_r\gtrsim\mathcal{O}(1)}$ a non-local mechanism contributes to the inward drift that generates the clustering through the statistical asymmetry of the path-history of approaching and separating particle pairs.  

The claim regarding the universality of the clustering mechanism across the range of scales in turbulence is in apparent disagreement with the sweep-stick mechanism put forth by Coleman \& Vassilicos \cite{coleman09}. However, we have shown that when ${St_r\ll1}$ in the inertial range, the sweep-stick mechanism is basically equivalent to our mechanism if the particles are suspended in Navier-Stokes turbulence. When ${St_r\gtrsim\mathcal{O}(1)}$ in the inertial range, the sweep-stick mechanism breaks down due to the increasing importance of the non-local clustering mechanism, which is not captured by the sweep-stick model.

Finally, we applied our results for the form of the drift velocity in the regime ${St_r\ll1}$ in the inertial range to the model equation for the RDF from \cite{zaichik07}.  Using this we obtained a prediction for the analytic form of the RDF in the inertial range when ${St_r\ll1}$. In contrast to the dissipation range, the RDF in the inertial range is not scale invariant, and this can be traced to the $r$ dependence of $\tau_r$ in the inertial range. Comparisons with DNS data demonstrated the accuracy of the prediction.\\

\noindent The authors acknowledge financial support from the National Science Foundation through Grant CBET-0967349 and through the Graduate Research Fellowship awarded to PJI.  Computational simulations were performed on Yellowstone \cite{yellowstone} (ark:/85065/d7wd3xhc) at the U.S. National Center for Atmospheric Research through its Computational and Information Systems Laboratory (sponsored by the National Science Foundation).

\bibliographystyle{unsrt}
\bibliography{refs_co12}

\end{document}